\documentclass{aa}
\usepackage{graphicx}
\usepackage{txfonts}
\usepackage{url}
\usepackage{natbib}
\usepackage{color}

\newcommand{\arcsecs}{\hbox{$^{\prime\prime}$}}

\begin{document} 
\title{Excitation of an Outflow From the Lower Solar Atmosphere \\ and a Co-Temporal EUV Transient Brightening}
\titlerunning{Blinker Formation}

\author{C.J.~Nelson\inst{1,2} \& 
	     J.G.~Doyle\inst{1}}

\institute{Armagh Observatory, College Hill, Armagh, BT61 9DG, UK
\and
School of Mathematics and Statistics, University of Sheffield, Hicks Building, Hounsfield Road, Sheffield, S3 7RH, UK \\
\email{c.j.nelson@shef.ac.uk}}

   \date{}

 \abstract{}
{We analyse an absorption event within the H$\alpha$ line wings, identified as a surge, and the co-spatial evolution of an EUV brightening, with spatial and temporal scales analogous to a small blinker.}
{We conduct a multi-wavelength, multi-instrument analysis using high-cadence, high-resolution data, 
collected by the {\it Interferometric BIdimensional Spectrometer} on the Dunn Solar Telescope, as 
well as the space-bourne {\it Atmospheric Imaging Assembly} and {\it Helioseismic and Magnetic 
Imager} instruments onboard the Solar Dynamics Observatory.}
{One large absorption event situated within the plage region trailing the lead sunspot of AR 11579 
is identified within the H$\alpha$ line wings. This event is found to be co-spatially linked to a 
medium-scale (around $4$\arcsecs\ in diameter) brightening within the transition region and corona. 
This ejection appears to have a parabolic evolution, first forming in the H$\alpha$ blue wing before 
fading and reappearing in the H$\alpha$ red wing, and comprises of a number of smaller fibril events. The line-of-sight photospheric magnetic field shows no evidence of cancellation leading to this event.}
{Our research has identified clear evidence that at least a subset of transient brightening events 
in the transition region are linked to the influx of cooler plasma from the lower solar atmosphere during 
large eruptive events, such as surges. These observations agree with previous numerical researches on the nature of blinkers and, therefore, suggest that magnetic reconnection is the driver of the analysed surge events; however, further research is required to confirm this.}{}

   \keywords{Sun: photosphere -  Sun: chromosphere -  Sun: transition region - Sun: corona}

   \maketitle

\section{Introduction}
The solar atmosphere is a dynamic environment which has been observed to ubiquitously change 
on ever smaller-scales with the advancement of observational techniques. The complex interactions 
between the strong, inferred magnetic fields and plasma flows are believed to lead to the formation 
of a number of structures which are easily identified from the photosphere to the corona ({\it e.g.}, 
spicules, macro-spicules, surges, coronal loops, prominences). Many such solar structures have been widely studied in recent years, due to their assumed influence in heating the corona to multi-million Kelvin temperatures.

Blinkers, small-scale (often between $3$\arcsecs\ and $10$\arcsecs\ in diameter) and short-lived 
(less than $20$ minutes) brightening events, have been extensively researched in the past two decades 
due to their dynamic nature. First observed with the {\it Coronal Diagnostic Spectrometer} 
(CDS; described by \citealt{Harrison95}) instrument, onboard the Solar and Heliospheric Observatory 
(SOHO), blinkers are often discussed with respect to chromospheric and coronal extreme UV (EUV) lines 
(see, for example, \citealt{Harrison97}, \citealt{Berghmans98}, \citealt{Doyle2004}). The nature of these 
transient brightenings in the EUV was analysed by \citet{Harrison99} who hypothesized that blinkers 
in the upper atmosphere were caused by an increase in density or filling factor, rather than due to a 
localized temperature enhancement (a result which has been supported by, {\it e.g.}, \citealt{Teriaca01}, 
\citealt{Bewsher03}, \citealt{Madjarska03}). 

Combining SOHO/CDS observations with magnetograms inferred using the {\it Michelson Doppler Imager} 
(MDI; \citealt{Scherrer95}) instrument, \citet{Bewsher02} and \citet{Parnell02} analysed the links 
between blinkers and the underlying photospheric magnetic field within the quiet Sun (QS) and active 
regions (AR), respectively. It was found that a high proportion of blinkers (approximately $75$\%) 
occurred co-spatially to strong uni-polar fields and, hence, it was inferred that magnetic reconnection 
within the upper atmosphere was unlikely to be the driver of these events.  These results agreed with the interpretations of \citet{Priest02}, who suggested that the forced interactions of cooler plasma from the lower solar atmosphere and warmer transition region plasma through a driver in the 
photosphere could lead to the formation of blinkers 
due to density enhancements. Indeed, five specific scenarios were discussed and a cartoon was 
presented which detailed each of the proposed mechanisms.

Within QS researches, it has often been reported that there is rarely a signature of blinker events 
within coronal lines (see, {\it e.g.}, \citealt{Harrison97}, \citealt{Bewsher02}, \citealt{Madjarska03}). 
A study of AR events, however, by \citet{Parnell02} found that a significant number of blinkers 
were co-spatial to intensity enhancements in coronal lines (such as the \ion{Mg}{IX} and \ion{Mg}{X} 
ions which sample plasma at approximately $10^6$). It was 
hypothesized that the increased intensity within these lines corresponded to the higher background 
emission of the corona within ARs and that, therefore, higher sensitivity within the QS may return 
similar results. Indeed, more recent work by \cite{Madjarska09} and \cite{Sri2012} showed that the term blinker 
covers a range of phenomena with blinkers originating at both chromospheric, transition region and coronal  
heights. 

\begin{figure*}
\includegraphics[scale=0.7]{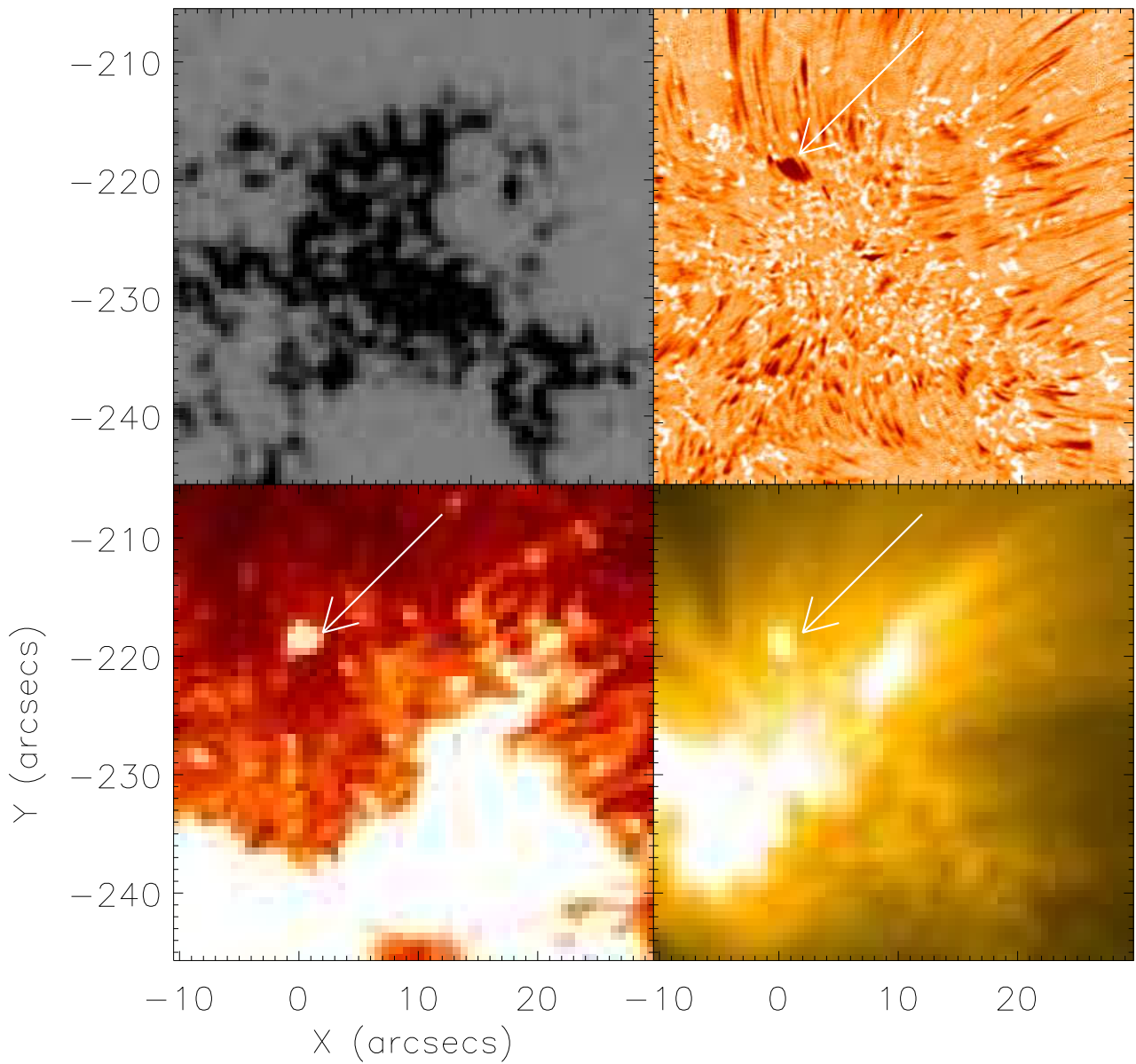}
\caption{Snapshot of data used in this study at 17:30:01 UT. (Top left) SDO/HMI, (Top right) H$\alpha$ 
blue wing, (Bottom left) SDO/AIA $304$ \AA, and (Bottom right) SDO/AIA $171$ \AA. The plotted 
arrow indicates the event which we discuss in this article.}
\label{Fig1}
\end{figure*}

One suggested mechanism (see \citealt{Priest02}) by which blinkers could be formed is by the ejection 
of cool plasma into the upper atmosphere (potentially manifesting as, {\it e.g.}, spicules). An example of a larger-scale  eruptive phenomena observed to propagate from the lower solar atmosphere into the transition region and corona are solar surges. These events have been observed for several decades (see, for example, \citealt{Roy73a, Roy73b} and references therein) and manifest as long, thin, dark jets in the H$\alpha$ line wings, often associated with transition region and coronal brightenings ({\it e.g.}, \citealt{Jiang07}, \citealt{Madjarska09}, \citealt{Kayshap13}). Due to the strong correlation between surge formation and strong, opposite polarity field regions, it has been widely suggested that these events are formed through magnetic reconnection in the lower atmosphere. \citet{Yokoyama96} performed numerical simulations of the solar atmosphere, finding evidence of both slow shocks and plasma ejections, analogous to surges, away from a magnetic reconnection site leading to simultaneous X-ray brightenings and H$\alpha$ surge flows.

\begin{figure*}
\includegraphics[scale=0.7]{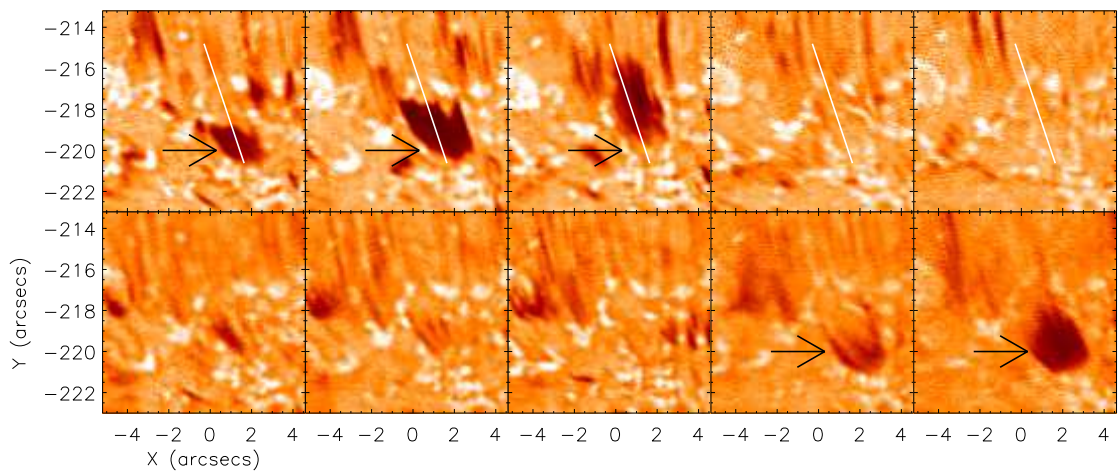}
\caption{Temporal evolution in the blue and red winsg of the H$\alpha$ line profile. (Top row) The evolution of this event through time (from 17:29:42 UT to 17:35:35 UT) in the  blue wing (each frame is separated by approximately $70$ seconds). (Bottom row) Corresponding 
FOV and frames in the red wing showing the apparent parabolic trajectory of this event. 
The white line indicates an artificial slit used to enable a time-distance plot (shown in Fig.~\ref{Fig3}).}
\label{Fig2}
\end{figure*}

Due to the many similarities in morphology, a concerted effort has been made in recent years to 
test whether eruptive events, such as surges or macro-spicules, and blinker phenomena, 
are linked. \citet{OShea05} analysed data at the limb and found events analogous to blinkers occurring co-spatially to regions of dynamic motions, including the ejection of cool photospheric plasma through spicules.  It was suggested that the most likely cause of blinkers was the heating of spicular mass within the upper-atmosphere. This work was continued by 
\citet{Madjarska06}, who studied lightcurves from two on disc blinkers and one limb event 
finding comparable evolutions through time. To fully understand this link, \citet{Priest02} 
suggested co-spatial and co-temporal H$\alpha$ and EUV observations (such as the \ion{He}{II} ion) 
would be required.

The results in this article are obtained using high-resolution, high-cadence data from both ground-based 
and space-borne instruments to analyse the formation of a single event in the H$\alpha$ line wings, 
which is co-spatial with a brightening at EUV wavelengths in both the transition region and the 
corona. We format our research as follows: In Section 2, we discuss the observations used in this 
article; Section 3 presents the results of our analysis before we conclude in Section 4. Finally, 
we make a short discussion of the implications of this study in Section 5.

\section{Observations}

In this article, co-temporal datasets from three instruments are analysed. We make use of 
ground-based data collected using the {\it Interferometric BIdimensional Spectrometer} 
(IBIS; \citealt{Cavallini00}) instrument situated at the National Solar Observatory, New Mexico. 
These data were collected during a period of good seeing on the $30^{\mathrm{th}}$ September, 
2012 between 17:22:02 UT and 17:37:06 UT and consist of $377$ speckled frames (see \citealt{Woger08}) 
in both the blue and red wings (approximately $\pm{0.75}$ \AA) of the H$\alpha$ line profile. 
IBIS was set to sample the plasma in a $97$\arcsecs\ by $97$\arcsecs\ field-of-view (FOV), 
trailing a large sunspot in AR 11579, which contained a small, dynamic pore. This FOV 
contained a large network structure co-spatial to strong, uni-polar magnetic fields and appeared 
to be relatively stable throughout these observations. The final, fully reduced cadence of 
these data is $2.4$ seconds, with pixel sizes and approximate spatial resolutions of $0.097$\arcsecs\ and 
$0.2$\arcsecs, respectively.

Data from the {\it Atmospheric Imaging Assembly} (AIA; as described by \citealt{Golub06}) onboard 
the Solar Dynamics Observatory are also studied to conduct a multi-wavelength analysis of the 
upper atmosphere. A number of wavelengths are analysed, from the $1700$ \AA\ photospheric continuum 
(used to align the instruments) to the transition region and coronal wavelengths. These data 
have a cadence of $24$ seconds (for the $1700$ \AA) and $12$ seconds, pixel sizes of $0.6$\arcsecs\ 
and, hence, a spatial resolution of around $1.5$\arcsecs. Each SDO/AIA wavelength is cropped and 
de-rotated to follow the same $97$\arcsecs\ by $97$\arcsecs\ FOV through time. 

To infer the topology of the under-lying magnetic field, a two-hour period (between 16:00:33 UT and 17:59:48 UT) of data from the {\it Helioseismic and Magnetic Imager} (HMI; \citealt{Graham03}) are also analysed. Over the period of these observations, 
the magnetic flux within the region studied in this article appeared stable and consisted of a 
negative polarity plage structure, which trailed a large, positive polarity sunspot. HMI data have 
a cadence of around $45$ seconds and pixel sizes and spatial resolutions of approximately 
$0.5$\arcsecs\ and $1$\arcsecs, respectively. In Fig.~\ref{Fig1}, we present a zoomed FOV of these 
data, including the background magnetic field from the HMI instrument, the H$\alpha$ blue wing, 
SDO/AIA $304$ \AA, and $171$ \AA\ filters. The event studied in detail in this article is indicated by the white arrow.

\section{Results}

\subsection{Evolution within the H$\alpha$ wings}

The event studied in this article is first observed in the H$\alpha$ blue wing as a medium-sized 
ejection emanating from the edge of a large region of network. In Fig.~\ref{Fig2}, we plot 
the evolution of this event through time for both the blue (top row) and the red (bottom row) 
wings of the H$\alpha$ line; each column images co-temporal frames between the wings.  
Note the apparent parabolic shape of the event, where the strong absorption firstly 
occurs in the blue wing before fading away and then occurring in the red wing. Although 
these data only cover a period of $15$ minutes, the full evolution of the event in the blue wing 
is observable, however, the absorption in the red wing still covers a significant area in the 
final frame. As the emission in the SDO/AIA lines returns to normal well before the end of these 
observations, we assume that the lack of information for the end of the lifetime of the event does 
not interfere with our conclusions.

The apparent morphology of this event, in the H$\alpha$ blue wing, shows rapid changes in both 
length and width over time. The event begins as a small number of fine threads, analogous to 
the near ubiquitous fibrils observed at the right of the FOV in Fig.~\ref{Fig1}, being emitted 
from similar spatial positions before forming a small `blob', around $1$\arcsecs\ by $1$\arcsecs\ 
in area, as seen in the initial frame of Fig.~\ref{Fig2}. This `blob' then expands in width and 
length to nearly $4$\arcsecs\ by $4$\arcsecs\ at its peak, before the absorption fades. 
Interestingly, once the event reaches a peak length, it propagates away from the footpoint, as seen in 
the third frame of Fig.~\ref{Fig2}, suggesting an ejection of a finite amount of mass evacuated 
away from the driver over time. The white line over-plotted on the blue wing images indicates 
the axis parallel to the direction of propagation analysed in this article.

\begin{figure}
\includegraphics[scale=0.5]{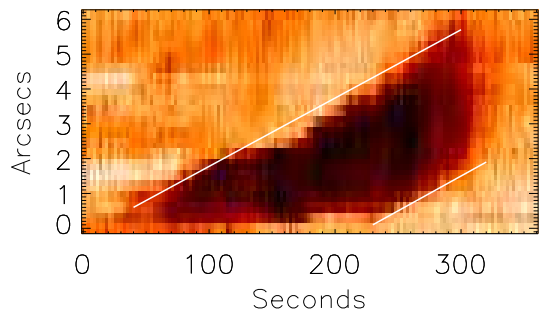}
\caption{Time-distance plot showing the apparent horizontal velocity of this event. White lines 
are over-plotted and are used to calculate the gradient of the motion, and hence the velocity 
within the H$\alpha$ blue wing. The slit position is shown in Fig.~\ref{Fig2}.}
\label{Fig3}
\end{figure}

In Fig.~\ref{Fig3}, we plot a time-distance diagram for a slit taken parallel to the axis of the 
event. Over time, the absorption propagates outwards from a footpoint at a steady speed (indicated 
by the white line), estimated to be around $14$ km s$^{-1}$, which is close to the sound speed 
for the photosphere and lower chromosphere. It is interesting to note that as the absorption feature fades away 
from the footpoint, approximately the same apparent velocity can be measured. The vertical 
component of the propagation would also be required to calculate an accurate total velocity. A 
large increase in the actual velocities compared to the apparent velocities is found if 
a significant vertical component of motion exists, by assuming a constant direction of 
propagation parallel to the axis of the event as observed in the H$\alpha$ blue wing. For 
example, a $30^{\circ}$ angle of propagation would be required for an increase in velocity to 
$16$ km s$^{-1}$. Larger angles such as $60^{\circ}$ or $85^\circ$ would lead to velocities of 
$28$ km s$^{-1}$ and $163$ km s$^{-1}$, respectively. We note that 3-D data would be required to 
accurately calculate such a velocity, however, a brief search indicated that no Solar Terrestrial 
Relations Observatory (STEREO) co-spatial data is available and that, therefore, we are unable 
to make any bolder assertions.

\begin{figure}
\includegraphics[scale=0.23]{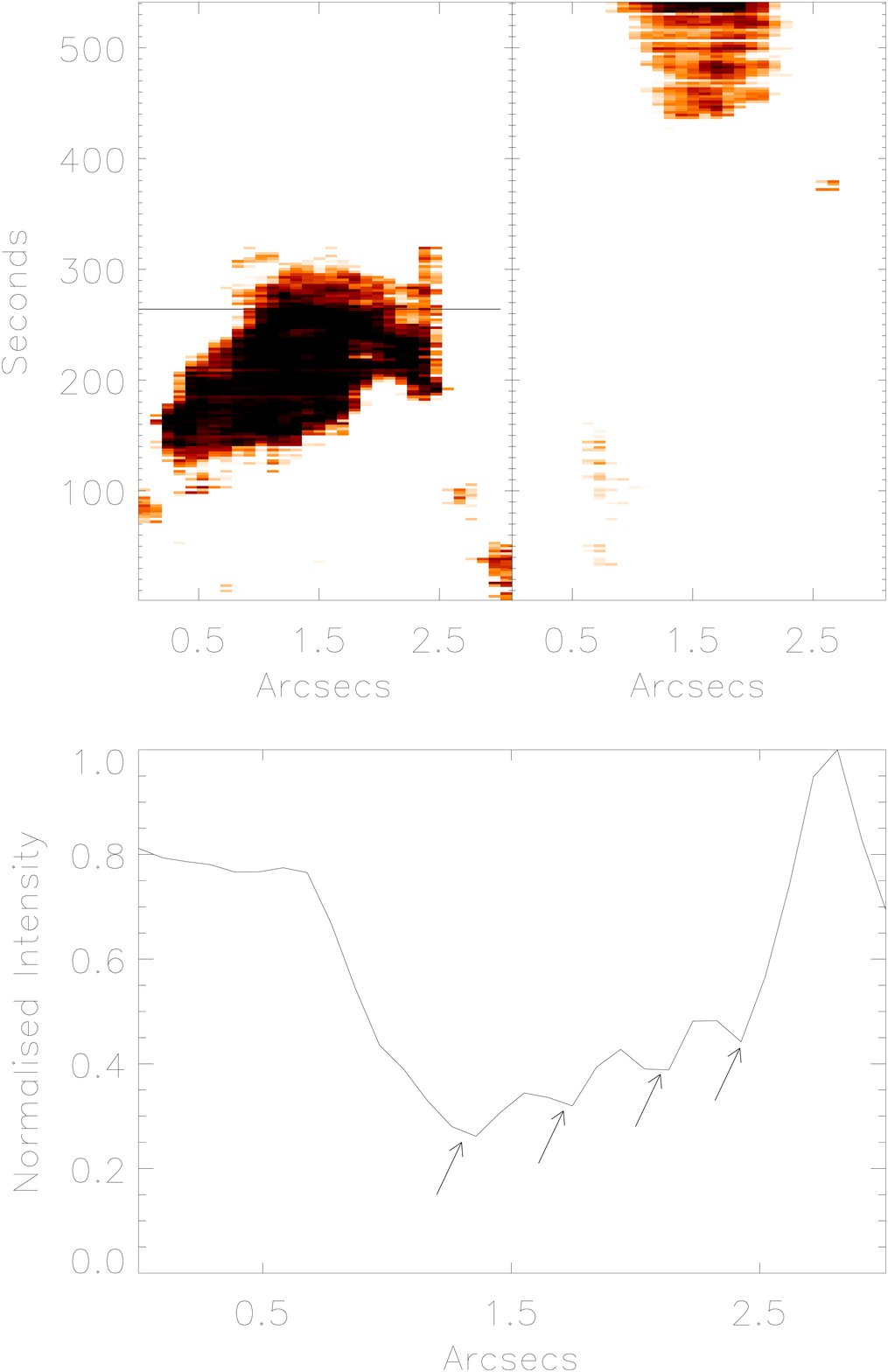}
\caption{Emission of the H$\alpha$ line wings for a slit perpendicular to the event, highlighting the small scale structuring within the events. (Top row) Time-distance plots taken for a slit across the event in the blue wing 
(left) and red wing (right). We emphasize the threads in this plot by enhancing the contrasts 
of the image. A black slit is over-plotted to indicate the region plotted with respect to 
intensity (bottom row). Note, the quadruple intensity minima between $1$\arcsecs\ and 
$2.5$\arcsecs\ indicated by the arrows.}
\label{Fig4}
\end{figure}

\subsection{Small-scale structuring}

Throughout the lifetime of the event as depicted in Fig.~\ref{Fig2} (the full temporal evolution is available as a movie in the online edition), small thread-like structures can 
be observed within the larger `blob'. These threads are reminiscent of the common fibrils which 
are easily seen in Fig.~\ref{Fig1}, however, their apparent length is shorter. Unfortunately, 
it is impossible to infer whether the difference in length is a morphological trait or a 
line-of-sight issue and, therefore, we do not discuss it further. In the second and third 
frames of Fig.~\ref{Fig2}, several of these thin structures are evident within the larger 
structure. In Fig.~\ref{Fig4}, time-distance plots taken for a co-spatial slit in the blue 
(left) and red (right) wings are presented showing the existence of small-scale structures 
across the event. Each time-distance diagram is calculated by analysing a slit perpendicular 
to the propagation of the `blob' through time. From these plots, it is simple to calculate 
that the absorption between the wings is separated by approximately $100$ seconds; we suggest 
that this gap indicates the position of the peak height, where the mass slows, and then falls.

\begin{figure*}
\includegraphics[scale=0.65]{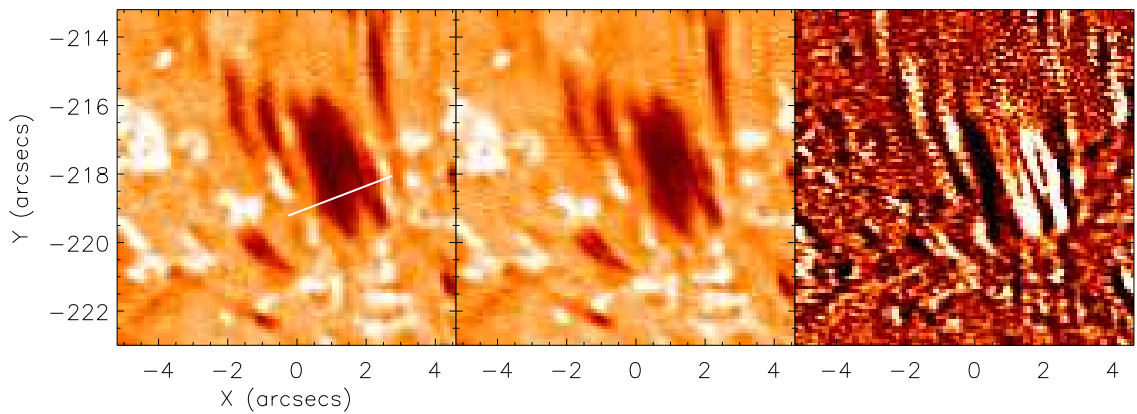}
\caption{Evolution of the event through time highlighting the fine structured nature of this event. (Left frame) H$\alpha$ blue wing emission at 17:32:25 UT ; the white line indicates the slit selected for analysis in Fig.~\ref{Fig4}. (Centre frame) The H$\alpha$ blue wing emission at 17:32:27 UT ({\it i.e.}, the consecutive frame). 
(Right frame) Difference image between the frames highlighting the fine structures within this event.}
\label{Fig5}
\end{figure*}

A black line in the blue wing plot of Fig.~\ref{Fig4} indicates the time frame used in the 
final (bottom) frame of Fig.~\ref{Fig4}. A number of rapid intensity changes are observed to 
exist along the slit, which are evidence of fine-scale structuring within the larger event. 
Between $1$\arcsecs\ and $2.5$\arcsecs, four minima can be found (indicated with arrows), 
each with widths of approximately $0.4$\arcsecs, consistent with other fibril structures 
within the feature. This small-scale structuring is similar to evidence presented 
during observations of surges (see, for example, \citealt{Lin2008}), where 
a number of thin-threads are observed within the larger event. We suggest that the absorption 
structure researched within this article is evidence of an on disc surge, comprising of a number of 
smaller-scale ejections, originating from a strong uni-polar plage region.

To further highlight these smaller structures, difference images were analysed frame-by-frame. 
We implemented a running difference technique, whereby $Diff[x,y,t+1]=Frame[x,y,t+1]-Frame[x,y,t]$, 
to identify changes in intensity between consecutive frames. In Fig.~\ref{Fig5}, we plot a 
representative example of the output of such an analysis, including the image analysed in the 
bottom frame of Fig.~\ref{Fig4}, the consecutive frame, and the returned difference image. The 
white line plotted in the image on the left indicates the slit analysed by Fig.~\ref{Fig4}. 
Within the imaging data, it is difficult to see the rapid changes in intensity between the 
individual fibril events, however, the difference image clearly shows the existence of the 
small scale structures. Each of these structures appears to evolve in a similar manner to 
normal fibrils, showing a marked increase in length (as shown in Fig.~\ref{Fig3}) and a 
parabolic shape. Such evolutions have been widely observed within Type I spicules and agree with current theories on $p$-mode buffeting (\citealt{dePontieu04}) and 
magnetic reconnection (\citealt{Yokoyama96}).

\subsection{Links to the photospheric magnetic field}

The relationship between surges and strong photospheric magnetic fields has been well documented in recent years (see, for example, \citealt{Roy73b}, \citealt{Madjarska09}, \citealt{Kayshap13}). Often, bright features identified as Ellerman bombs (EBs; first reported by \citealt{Ellerman17}) are observed at the footpoints of surges and have been interpreted as evidence of the dynamic nature of the magnetic field co-spatial to the apparent footpoint of these events (for example, \citealt{Roy73a}). More recently, links between surges and flux cancellation within an AR have been presented by \citet{Chae99} and \citet{Chen09} who suggested that cancellation was evidence supporting reconnection as a driver for the ejection of plasma from the lower chromosphere.

The surge identified in Sections 3.1 and 3.2 is observed to form co-spatially to a large uni-polar plage region (as is easily seen in the first frame of Fig.~\ref{Fig1}). The magnetic field within this FOV appears to be stable throughout the course of these data, showing no evidence of flux emergence or cancellation. As has been discussed, the majority of surges analysed by previous researches have found evidence of significant dynamics within co-spatial magnetic fields, however, our results show no such activity. Despite small-scale restructuring of the magnetic field within the plage region being evident co-temporally with the beginning of the absorption feature, it is difficult to quantify how much influence, if any, this has on the ejection.

We, therefore, further analyse the magnetic evolution of this FOV over the course of a two-hour period surrounding the data discussed in this article. This set of data confirms that no bi-polar fields interact to form this event and that only small-scale restructuring occurs within the large uni-polar plage region. This result means that we are unable to present any evidence of magnetic cancellation, or reconnection, leading to the formation of this event. It should be noted that although there are few dynamic changes within the FOV, significant topological complexities of the magnetic field can be inferred through the number and frequency of network bright points, which are often used as a proxy for the vertical magnetic field (see, {\it e.g.}, \citealt{Leenaarts06}). It is, therefore, possible that magnetic structuring on scales smaller than are currently resolvable using the HMI instrument is occuring during this period and leading to reconnection of complex topologies, but the set of data analysed here shows no evidence of this.

\begin{figure}
\includegraphics[scale=0.35]{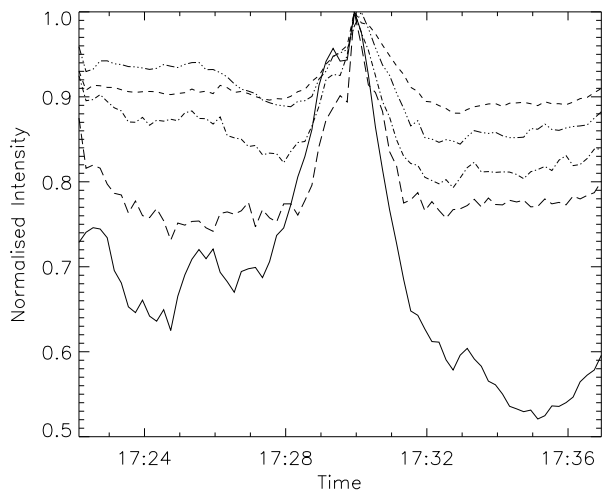}
\caption{Lightcurves showing the evolution of the upper-atmosphere during the course of this 
event. Plotted, from bottom-to-top in the first frame, are the SDO/AIA $304$ \AA\ (solid line), 
$131$ \AA\ (long-dashed), $211$ \AA\ (dot-dashed), $171$ \AA\ (short-dashed), and $193$ \AA\ 
(dot-dot-dot-dashed) filters.}
\label{Fig6}
\end{figure}

\subsection{Signatures in the transition region and corona}

Finally, we assess the effect this event has on the upper atmosphere through an analysis 
of the SDO/AIA EUV filters. It can be easily inferred from Fig.~\ref{Fig1}, that a co-spatial 
brightening event can be seen within the AIA $304$ \AA\ and $171$ \AA\ filters during a period of 
strong absorption within the H$\alpha$ blue wing. Within the $304$ \AA\ filter, the peak area 
of this brightening is approximately $4$\arcsecs\ by $4$\arcsecs, consistent with the area 
of a small blinker event (as discussed by, {\it e.g.}, \citealt{Chae00}, \citealt{Parnell02}, 
\citealt{Madjarska03}). We note, however, that the spatial resolution of the SDO/AIA 
instrument is significantly higher than the SOHO/CDS instrument used by those researches and, 
hence, that the scale of this event may appear smaller due to instrumental improvements and 
not a physical difference.

The evolution of the area of this brightening within the $304$ \AA\ filter is roughly parabolic, 
showing a steady rise to the peak area, before dropping off once again to the background 
intensity. This parabolic evolution is also apparent within all other SDO/AIA upper atmospheric 
filters co-temporally, implying that these brightenings are formed as a result of increased 
density or filling factor in the transition region and corona, rather than a localized 
heating event, as has been widely suggested in previous researches (see, {\it e.g.}, 
\citealt{Priest02}). We suggest that the large absorption feature observed within the 
H$\alpha$ blue wing is linked to the supply of mass from the lower atmosphere into the upper atmosphere. This 
presents the open question which is not answered by this article; do smaller fibril structures 
which are observed within the line wings have a similar, albeit smaller, influence on the 
upper atmosphere?

The temporal evolution of this event is plotted in Fig.~\ref{Fig6} for a number of SDO/AIA 
filters. A $7$\arcsecs\ by $6$\arcsecs\ box was selected around this apparent 
small-scale blinker such that no other localized brightening events occurred within the FOV 
during this period. The average intensity of this box was then calculated for each frame 
during these observations and plotted for each of five EUV SDO/AIA filters (from bottom to 
top in the original frame: $304$ \AA; $131$ \AA; $211$ \AA; $171$ \AA; and $193$ \AA). 
The localized brightening event occurs for approximately four minutes which is, once again, 
the lower boundary for lifetimes of blinkers. The improved cadence of the SDO/AIA instrument 
over the SOHO/CDS instrument may explain this. 

\begin{figure*}
\includegraphics[scale=0.75]{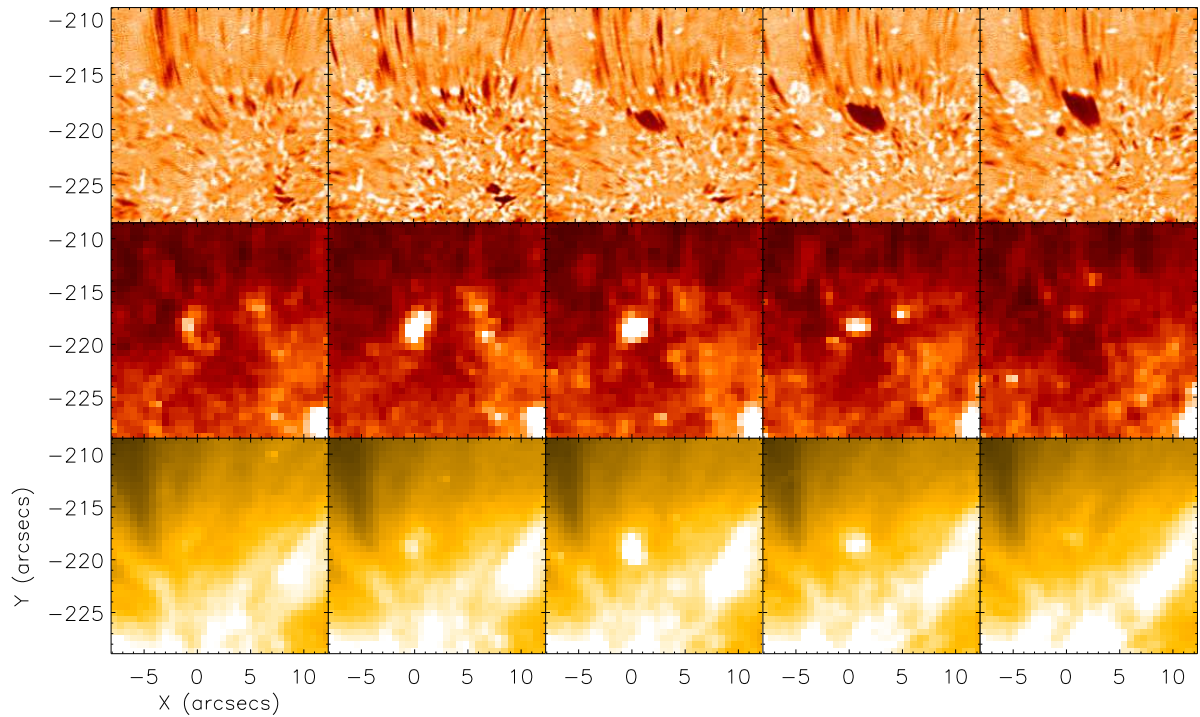}
\caption{The evolution of the event analysed in this article in the H$\alpha$ blue wing (top row), 
SDO/AIA $304$ \AA\ (middle row), and $171$ \AA\ (bottom row) filters. The difference 
between each frame is approximately $48$ seconds (starting at 17:28:26 UT and ending at 17:31:37 UT); 
we estimate that the temporal difference between each wavelength is below $10$ seconds.}
\label{Fig7}
\end{figure*}

It is interesting to note, that the original brightening in the SDO/AIA filters occurs co-temporally 
with the initial stages of the formation of the H$\alpha$ event (similar to the numerical results 
of \citealt{Yokoyama96}), before fading entirely whilst 
there is still strong absorption in the blue wing. In Fig.~\ref{Fig7}, we plot a visualization 
of this process. The second frame of Fig.~\ref{Fig7} highlights the delay of absorption in the 
H$\alpha$ blue wing perfectly. Only the initial stages of development within the H$\alpha$ blue 
wing are observed in this frame, but near peak emission in the SDO/AIA $304$ \AA\ data are 
inferred. One possible reason for the time lag within these data could be that we only observe 
one specific spectral position in each wing within the H$\alpha$ line and that, therefore, we 
only receive a snapshot of the whole physical process which is occurring. We note that if full 
H$\alpha$ line scan data were available to analyse during this event, a better picture of the 
observed coupling may have been inferred. We do, however, suggest that this strong 
absorption feature in the H$\alpha$ line wing is intrinsically linked to the possible blinker 
event observed within the SDO/AIA images.

We suggest that the brightening observed within these SDO/AIA data is analogous to a blinker. 
Although the spatial and temporal scales of this event are on the lower limit of previously 
observed blinkers (such as those observed by, for example, \citealt{Harrison97}, 
\citealt{Harrison99}, \citealt{Bewsher02}), the improved spatial and temporal resolutions 
of the data used within this study may account for this. Combining this result with the large 
absorption feature within the H$\alpha$ line wings, interpreted as an on disc surge, 
presents additional evidence that at least a subset of transient 
brightening features within the transition region and corona are linked to mass 
supply from the lower atmosphere. Due to the co-temporal reaction of the H$\alpha$ line wings and the EUV filters, we suspect that the increased filling factor in the upper atmosphere is caused by a slow shock, propagating away from a reconnection site in the lower atmosphere (in a way described by \citealt{Yokoyama96}). 

\section{Discussion} 
In this article, we have presented a single large-scale absorption feature which is co-temporal 
to an increased emission from the SDO/AIA transition region and coronal filters. We find strong 
evidence implying that the absorption feature in the H$\alpha$ line wings is analogous to jet events, such as
surges and macro-spicules, as observed at limb, as well as being linked to the potential small-scale 
blinker event within the corona.

A question to be addressed is the blinker formation mechanism. 
\citet{Doyle2004} discussed blinker phenomena being associated with brightenings in pre-existing 
coronal loops. The brightenings appeared to occur during the emergence of new magnetic flux.
These authors suggested that the blinker activity was triggered by interchange reconnection, 
serving to provide topological connectivity between newly emerging flux and pre-existing flux 
with the {\it EUV Imaging Telescope} (EIT) images showing the existence of loop structures prior to 
the onset of the blinker activity. The temperature interfaces created in the reconnection process 
between the cool plasma of the newly emerging loop and the hot plasma of the existing loop was 
suggested to cause the observed activity seen in both the SUMER and CDS data. As the temperature 
interfaces propagate with the 
characteristic speed of a conduction front they heat up the cool chromospheric plasma to coronal 
temperatures and increase the volume which brightens to transition region temperatures. 
This was followed-up by \citet{Sri2008} who analysed several blinkers and found that in most 
instances they were associated with the emergence of magnetic flux which gave rise to the appearance 
of multiple magnetic reconnection events, across an upper atmosphere (coronal) magnetic 
null point, along with a loop structure as observed with the TRACE 171 \AA\ filter. In a more recent 
study, \citet{Sri2012} classified blinkers into two categories, one associated with coronal 
counterparts and other with no coronal counterparts as seen in XRT images and EIS 195 \AA\ raster 
images. Around two-thirds of the blinkers show coronal counterparts and correspond to various events 
like EUV/X-ray jets, brightenings in coronal bright points or foot-point brightenings of larger 
loops. The present event fits into those showing a coronal component and matches the slow-shock model presented by \citet{Yokoyama96} suggesting that magnetic reconnection may be the driver of this event.

It should be noted, however, that the lower atmosphere shows no evidence supporting the occurrence of magentic reconnection. Due to the lack of activity within the observed vertical magnetic field, and the structured nature of the absorption event, magnetic reconnection is potentially not the driver of this surge. It is, however, possible that magnetic reconnection is occuring within the lower solar atmosphere, in a manner similar to Ellerman bombs (EBs; see, {\it e.g.}, \citealt{Ellerman17}, \citealt{Nelson13}). Both \citet{Lee03} and \citet{Kayshap13} found evidence of energy release in the upper solar atmosphere following potential reconnection events in the photosphere. To fully test whether this is the mechanism observed here, we would require a larger statistical sample of surges occuring around uni-polar regions.

\begin{acknowledgements}
Research at the Armagh Observatory is grant-aided by the N. Ireland Dept. of Culture, Arts 
and Leisure. We thank the National Solar Observatory / Sacramento Peak for their hospitality 
and in particular Doug Gilliam for his excellent help during the observations. IBIS data 
reductions were accomplished with help from Kevin Reardon. We thank the UK Science and 
Technology Facilities Council for CJN's studentship, PATT T\&S support, plus support from 
STFC grant ST/J001082/1 and the Leverhulme Trust. HMI data is courtesy of SDO (NASA) and 
the HMI consortium. We would also like to thank Friedrich W\"oger for his image reconstruction code.
\end{acknowledgements}

\bibliographystyle{aa}
\bibliography{Halphatocorona}

\begin{thebibliography}{35}
\expandafter\ifx\csname natexlab\endcsname\relax\def\natexlab#1{#1}\fi

\bibitem[{{Berghmans} {et~al.}(1998){Berghmans}, {Clette}, \&
  {Moses}}]{Berghmans98}
{Berghmans}, D., {Clette}, F., \& {Moses}, D. 1998, \aap, 336, 1039

\bibitem[{{Bewsher} {et~al.}(2002){Bewsher}, {Parnell}, {Brown}, \&
  {Hood}}]{Bewsher02}
{Bewsher}, D., {Parnell}, C.~E., {Brown}, D.~S., \& {Hood}, A.~W. 2002, in ESA
  Special Publication, Vol. 505, SOLMAG 2002. Proceedings of the Magnetic
  Coupling of the Solar Atmosphere Euroconference, ed. H.~{Sawaya-Lacoste},
  239--242

\bibitem[{{Bewsher} {et~al.}(2003){Bewsher}, {Parnell}, {Pike}, \&
  {Harrison}}]{Bewsher03}
{Bewsher}, D., {Parnell}, C.~E., {Pike}, C.~D., \& {Harrison}, R.~A. 2003,
  \solphys, 215, 217

\bibitem[{{Cavallini} {et~al.}(2000){Cavallini}, {Berrilli}, {Cantarano}, \&
  {Egidi}}]{Cavallini00}
{Cavallini}, F., {Berrilli}, F., {Cantarano}, S., \& {Egidi}, A. 2000, in ESA
  Special Publication, Vol. 463, The Solar Cycle and Terrestrial Climate, Solar
  and Space weather, ed. A.~{Wilson}, 607

\bibitem[{{Chae} {et~al.}(1999){Chae}, {Qiu}, {Wang}, \& {Goode}}]{Chae99}
{Chae}, J., {Qiu}, J., {Wang}, H., \& {Goode}, P.~R. 1999, \apjl, 513, L75

\bibitem[{{Chae} {et~al.}(2000){Chae}, {Wang}, {Goode}, {Fludra}, \&
  {Sch{\"u}hle}}]{Chae00}
{Chae}, J., {Wang}, H., {Goode}, P.~R., {Fludra}, A., \& {Sch{\"u}hle}, U.
  2000, \apjl, 528, L119

\bibitem[{{Chen} {et~al.}(2009){Chen}, {Jiang}, \& {Ma}}]{Chen09}
{Chen}, H., {Jiang}, Y., \& {Ma}, S. 2009, \solphys, 255, 79

\bibitem[{{De Pontieu} {et~al.}(2004){De Pontieu}, {Erd{\'e}lyi}, \&
  {James}}]{dePontieu04}
{De Pontieu}, B., {Erd{\'e}lyi}, R., \& {James}, S.~P. 2004, \nat, 430, 536

\bibitem[{{Doyle} {et~al.}(2004){Doyle}, {Roussev}, \& {Madjarska}}]{Doyle2004}
{Doyle}, J.~G., {Roussev}, I.~I., \& {Madjarska}, M.~S. 2004, \aap, 418, L9

\bibitem[{{Ellerman}(1917)}]{Ellerman17}
{Ellerman}, F. 1917, \apj, 46, 298

\bibitem[{{Golub}(2006)}]{Golub06}
{Golub}, L. 2006, \ssr, 124, 23

\bibitem[{{Graham} {et~al.}(2003){Graham}, {Norton}, {L{\'o}pez Ariste},
  {Lites}, {Socas-Navarro}, \& {Tomczyk}}]{Graham03}
{Graham}, J.~D., {Norton}, A., {L{\'o}pez Ariste}, A., {et~al.} 2003, in
  Astronomical Society of the Pacific Conference Series, Vol. 307, Solar
  Polarization, ed. J.~{Trujillo-Bueno} \& J.~{Sanchez Almeida}, 131

\bibitem[{{Harrison}(1997)}]{Harrison97}
{Harrison}, R.~A. 1997, \solphys, 175, 467

\bibitem[{{Harrison} {et~al.}(1999){Harrison}, {Lang}, {Brooks}, \&
  {Innes}}]{Harrison99}
{Harrison}, R.~A., {Lang}, J., {Brooks}, D.~H., \& {Innes}, D.~E. 1999, \aap,
  351, 1115

\bibitem[{{Harrison} {et~al.}(1995){Harrison}, {Sawyer}, {Carter}, {Cruise},
  {Cutler}, {Fludra}, {Hayes}, {Kent}, {Lang}, {Parker}, {Payne}, {Pike},
  {Peskett}, {Richards}, {Gulhane}, {Norman}, {Breeveld}, {Breeveld}, {Al
  Janabi}, {McCalden}, {Parkinson}, {Self}, {Thomas}, {Poland}, {Thomas},
  {Thompson}, {Kjeldseth-Moe}, {Brekke}, {Karud}, {Maltby}, {Aschenbach},
  {Br{\"a}uninger}, {K{\"u}hne}, {Hollandt}, {Siegmund}, {Huber}, {Gabriel},
  {Mason}, \& {Bromage}}]{Harrison95}
{Harrison}, R.~A., {Sawyer}, E.~C., {Carter}, M.~K., {et~al.} 1995, \solphys,
  162, 233

\bibitem[{{Jiang} {et~al.}(2007){Jiang}, {Chen}, {Li}, {Shen}, \&
  {Yang}}]{Jiang07}
{Jiang}, Y.~C., {Chen}, H.~D., {Li}, K.~J., {Shen}, Y.~D., \& {Yang}, L.~H.
  2007, \aap, 469, 331

\bibitem[{{Kayshap} {et~al.}(2013){Kayshap}, {Srivastava}, \&
  {Murawski}}]{Kayshap13}
{Kayshap}, P., {Srivastava}, A.~K., \& {Murawski}, K. 2013, \apj, 763, 24

\bibitem[{{Lee} {et~al.}(2003){Lee}, {Gallagher}, {Gary}, {Nita}, {Choe},
  {Bong}, \& {Yun}}]{Lee03}
{Lee}, J., {Gallagher}, P.~T., {Gary}, D.~E., {et~al.} 2003, \apj, 585, 524

\bibitem[{{Leenaarts} {et~al.}(2006){Leenaarts}, {Rutten}, {S{\"u}tterlin},
  {Carlsson}, \& {Uitenbroek}}]{Leenaarts06}
{Leenaarts}, J., {Rutten}, R.~J., {S{\"u}tterlin}, P., {Carlsson}, M., \&
  {Uitenbroek}, H. 2006, \aap, 449, 1209

\bibitem[{{Lin} {et~al.}(2008){Lin}, {Martin}, {Engvold}, {Rouppe van der
  Voort}, \& {van Noort}}]{Lin2008}
{Lin}, Y., {Martin}, S.~F., {Engvold}, O., {Rouppe van der Voort}, L.~H.~M., \&
  {van Noort}, M. 2008, Advances in Space Research, 42, 803

\bibitem[{{Madjarska} \& {Doyle}(2003)}]{Madjarska03}
{Madjarska}, M.~S. \& {Doyle}, J.~G. 2003, \aap, 403, 731

\bibitem[{{Madjarska} {et~al.}(2009){Madjarska}, {Doyle}, \& {de
  Pontieu}}]{Madjarska09}
{Madjarska}, M.~S., {Doyle}, J.~G., \& {de Pontieu}, B. 2009, \apj, 701, 253

\bibitem[{{Madjarska} {et~al.}(2006){Madjarska}, {Doyle}, {Hochedez}, \&
  {Theissen}}]{Madjarska06}
{Madjarska}, M.~S., {Doyle}, J.~G., {Hochedez}, J.-F., \& {Theissen}, A. 2006,
  \aap, 452, L11

\bibitem[{{Nelson} {et~al.}(2013){Nelson}, {Doyle}, {Erd{\'e}lyi}, {Huang},
  {Madjarska}, {Mathioudakis}, {Mumford}, \& {Reardon}}]{Nelson13}
{Nelson}, C.~J., {Doyle}, J.~G., {Erd{\'e}lyi}, R., {et~al.} 2013, \solphys,
  283, 307

\bibitem[{{O'Shea} {et~al.}(2005){O'Shea}, {Banerjee}, \& {Doyle}}]{OShea05}
{O'Shea}, E., {Banerjee}, D., \& {Doyle}, J.~G. 2005, \aap, 436, L43

\bibitem[{{Parnell} {et~al.}(2002){Parnell}, {Bewsher}, \&
  {Harrison}}]{Parnell02}
{Parnell}, C.~E., {Bewsher}, D., \& {Harrison}, R.~A. 2002, \solphys, 206, 249

\bibitem[{{Priest} {et~al.}(2002){Priest}, {Hood}, \& {Bewsher}}]{Priest02}
{Priest}, E.~R., {Hood}, A.~W., \& {Bewsher}, D. 2002, \solphys, 205, 249

\bibitem[{{Roy}(1973{\natexlab{a}})}]{Roy73b}
{Roy}, J.~R. 1973{\natexlab{a}}, \solphys, 32, 139

\bibitem[{{Roy}(1973{\natexlab{b}})}]{Roy73a}
{Roy}, J.~R. 1973{\natexlab{b}}, \solphys, 28, 95

\bibitem[{{Scherrer} {et~al.}(1995){Scherrer}, {Bogart}, {Bush}, {Hoeksema},
  {Kosovichev}, {Schou}, {Rosenberg}, {Springer}, {Tarbell}, {Title},
  {Wolfson}, {Zayer}, \& {MDI Engineering Team}}]{Scherrer95}
{Scherrer}, P.~H., {Bogart}, R.~S., {Bush}, R.~I., {et~al.} 1995, \solphys,
  162, 129

\bibitem[{{Subramanian} {et~al.}(2012){Subramanian}, {Madjarska}, {Doyle}, \&
  {Bewsher}}]{Sri2012}
{Subramanian}, S., {Madjarska}, M.~S., {Doyle}, J.~G., \& {Bewsher}, D. 2012,
  \aap, 538, A50

\bibitem[{{Subramanian} {et~al.}(2008){Subramanian}, {Madjarska}, {Maclean},
  {Doyle}, \& {Bewsher}}]{Sri2008}
{Subramanian}, S., {Madjarska}, M.~S., {Maclean}, R.~C., {Doyle}, J.~G., \&
  {Bewsher}, D. 2008, \aap, 488, 323

\bibitem[{Teriaca {et~al.}(2001)Teriaca, Madjarska, \& Doyle}]{Teriaca01}
Teriaca, L., Madjarska, M., \& Doyle, J. 2001, Solar Physics, 200, 91

\bibitem[{{W{\"o}ger} {et~al.}(2008){W{\"o}ger}, {von der L{\"u}he}, \&
  {Reardon}}]{Woger08}
{W{\"o}ger}, F., {von der L{\"u}he}, O., \& {Reardon}, K. 2008, \aap, 488, 375

\bibitem[{{Yokoyama} \& {Shibata}(1996)}]{Yokoyama96}
{Yokoyama}, T. \& {Shibata}, K. 1996, \pasj, 48, 353

\end{thebibliography}

\end{document}